\documentclass[pra,aps,twocolumn,showpacs,groupedaddress,superscriptaddress]{revtex4}
\usepackage{graphicx}

\begin{document}

\title{Dynamic optical bistability in resonantly enhanced Raman generation}

\author{I.\ Novikova}
\address{Harvard-Smithsonian Center for Astrophysics, Cambridge,
Massachusetts, 02138}
\author{A.\ S.\ Zibrov}
\address{Harvard-Smithsonian Center for Astrophysics, Cambridge,
Massachusetts, 02138}
\address{Department of Physics, Harvard University, Cambridge, MA 02138}
\address{Lebedev Institute of Physics, Moscow, 117924, Russia }
\author{D.\ F.\ Phillips} 
\address{Harvard-Smithsonian Center for Astrophysics, Cambridge,
Massachusetts, 02138}
\author{A.\ Andr\'e}
\address{Department of Physics, Harvard University, Cambridge, MA 02138}
\author{R.\ L.\ Walsworth}
\address{Harvard-Smithsonian Center for Astrophysics, Cambridge,
Massachusetts, 02138}
\address{Department of Physics, Harvard University, Cambridge, MA 02138}

\begin{abstract}
We report observations of novel dynamic behavior in 
resonantly-enhanced stimulated Raman scattering in 
Rb vapor.  In particular, we demonstrate a dynamic 
hysteresis of the Raman scattered optical field in 
response to changes of the drive laser field intensity 
and/or frequency.  This effect may be described as 
a dynamic form of optical bistability resulting from the 
formation and decay of atomic coherence.  
We have applied this phenomenon to the realization of an all-optical switch.
\end{abstract}


\date{\today}
\maketitle

\begin{figure} 
\centering
\includegraphics[width=1.0\columnwidth]{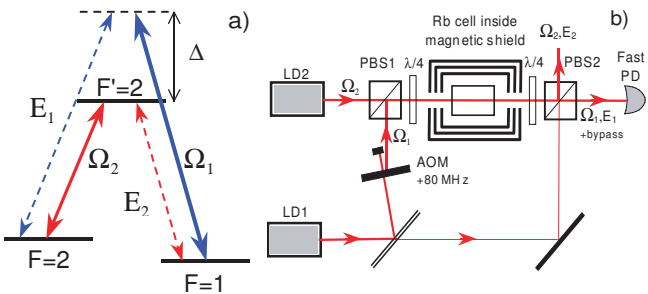}
\caption{a) Three-level double-$\Lambda$ scheme used in the experiment. The
first diode laser LD$1$ (pump field $\Omega_1$) is detuned by
$\Delta \approx 4$~GHz to the blue side of the Rb transition
$F=1\rightarrow F'=2$. The second laser LD$2$ (pump field $\Omega_2$) is
resonant with $F=2\rightarrow F'=2$.
b) Experimental setup.
\label{setup.fig}}
\end{figure}

It is now well-known that the optical properties of atomic media may be dramatically
altered if the atoms are placed into an appropriate quantum superposition of states,
enabling, e.g., electromagnetically induced transparency (EIT) and lasing without
inversion~\cite{scullybook}. Such coherently-prepared media can also exhibit
extremely large nonlinearity at very low light
levels~\cite{imamoglu'97prl,harris'98prl,lukin'99prl,lukin'00prl,harris'03}. In
this paper we demonstrate a novel form of nonlinearity -- dynamic optical
bistability -- using resonantly enhanced Raman generation in warm Rb vapor~\cite{lukin'00}. The observed bistable behavior results from the formation of a
long-lived coherence in the atomic ensemble due to the strong light-atom
interaction provided by a double-$\Lambda$ interaction scheme, which can be
approximated as two strong pump fields $\Omega_{1,2}$ interacting with a
three-level atom as shown in Fig.~\ref{setup.fig}a. Strong Raman
gain is produced with this scheme, generating a pair of correlated Stokes
and anti-Stokes fields $\mathrm{E}_{1,2}$~\cite{ICAP,wal'03,kuzmich'03nature}. 
We find dynamic optical bistability in the
form of hysteresis in the response of either of the Raman fields to
sufficiently fast variation in the corresponding pump field.

Optical bistability has been extensively studied
 for several decades (see~\cite{ob_book, lugiato'84} for reviews),
 and observed in many systems.
 Most typically, optical bistability occurs with a nonlinear medium placed in
 an optical cavity, such that there is more than one stable condition of
output  light intensity from the cavity for a given intensity of
near-resonant input light, i.e.,  there is a hysteresis in the effective cavity
transmission. Optical bistability with  a cavity has been observed with various
two- and three-level  atomic systems serving as the nonlinear medium~\cite{gibbs'76,bonifacio'78,cecchi'82,arecchi'83,xiao'02pra,xiao'03prl}. %
Optical bistability has also been demonstrated without a cavity using
degenerate four-wave mixing in atomic vapor with two
counter-propagating laser beams~\cite{gauthier'90,ackemann'97}. In these latter
experiments, the bistability arises from a dependence of
phase-matching conditions on the magnitude of the output field~\cite{winful'80}.

The unique features of the present results are: (i) optical bistability arises due to the formation of long-lived atomic coherence, which in turn depends on the amplitudes and phases of all
four optical fields ($\Omega_{1,2}$ and $\mathrm{E}_{1,2}$); and (ii) bistability is observed in a co-propagating laser geometry without a cavity. The steady
state amplitudes of both the Stokes and anti-Stokes fields are self-adjusted by
the light-atom interaction, so that a generalized four-photon dark state is established that is decoupled from all optical fields. The generalized dark
state in the double-$\Lambda$ system is analogous to the dark state polariton
employed in ``stored light'' in the single-$\Lambda$ system~\cite{hau'01nature,phillips'01prl,zibrov'02prl}, with the dynamics controlled by three relevant timescales. First, there is the
characteristic time of atomic response to change in the pump fields, i.e., the time for atomic coherence to be created and modified. This time  
is given by the inverse bandwidth of the four-photon Raman process~\cite{ICAP,wal'03},
and may be quite short:
\begin{equation} \label{tau_ram}
\tau_R \propto \frac{\Delta}{|\Omega_1||\Omega_2|},
\end{equation}
where $\Delta$ is the detuning of the off-resonant pump field $\Omega_1$ (see
Fig.~\ref{setup.fig}a). In our experiment, $\tau_R < 1~\mu\mathrm{s}$ for
typical values of the pump fields $\Omega_{1,2}$ and $\Delta$.

The second relevant timescale characterizes the equilibration of the four-photon dark state, and hence the response time of the amplitudes and phases of the generated Raman fields. This time is determined by the optical pumping rate of the far-detuned pump field $\Omega_1$:
\begin{equation} \label{tau_s}
\tau_{S} \propto \frac{\Delta^2}{\gamma|\Omega_1|^2},
\end{equation}
where $\gamma$ is the relaxation rate of the excited state. For comparable values of the pump field powers, one has $\tau_S \simeq \frac{\Delta}{\gamma}\tau_R \gg \tau_R$; for our system,  $\tau_S\approx 30~\mu\mathrm{s}$.
This large difference in timescales 
%for the
%adiabatic variation and relaxation of atomic coherence 
enables dynamic optical
bistability when either of the pump fields is modulated on a timescale between
$\tau_S$ and $\tau_R$, because the generated Raman fields depend on the magnitude
and phase of the atomic coherence. Once the modulated pump field passes
above the threshold for Raman generation, atomic coherence is created. However, the intensity of the Raman field reaches its steady-state value with some delay $\tau_S$, effectively increasing the observed threshold pump power. Similarly, Raman generation continues at lower pump field
power  than the steady-state threshold when the pump field is reduced from its peak value.
%until all the atomic coherence is converted into Raman photons. 
As the
pump field is again increased, the hysteresis cycle is repeated.

The third relevant timescale is the atomic coherence lifetime, which for our system is limited by atomic diffusion in and out of the interaction region to be $T_2\sim 2~\mathrm{ms}\gg \tau_S$. Because atomic coherence provides a ``reservoir'' for Raman generation, dynamic optical bistability can be observed for very slow pump field modulation, down to a modulation period $\simeq T_2$.
%
%We also need to take into account that atomic coherence may be stored 
%in the atoms for much longer time and, therefore, affect the dynamics of generated
%Raman field even at slower pump modulation. The relaxation time $T_2$ for the
%ground-state atomic coherence can be as long as $T_2 \sim
%2~\mathrm{ms} \gg \tau_S$ and it is limited by
%atomic diffusion in and out of the interaction region. If the pump modulation period is slower than $\tau_S$ and is comparable with $T_2$, similar hysteresis cycle is observed.  

A schematic of our experimental setup is shown in
Fig.~\ref{setup.fig}b.
We used two diode lasers LD$1$ and LD$2$ operating at
$795$~nm and resonant with the $D_1$ line ($5~{}^2S_{1/2} \rightarrow
5~{}^2P_{1/2}$) of ${}^{87}$Rb.
The first laser was detuned by $\Delta$ from the $F=1\rightarrow F'=2$
transition, and the second laser was resonant with the
$F=2 \rightarrow F'=2$ transition.
To control the intensity of the off-resonant field $\Omega_1$, a part of the
beam from the first laser was depleted using an acousto-optic modulator (AOM)
with a frequency shift of $80$~MHz, and then combined with radiation from the
second laser on a polarizing beam splitter (PBS$1$). The
polarizations of the two fields were then transformed by a quarter-wave plate into
orthogonal circular polarizations before
entering the Rb cell. The maximum available optical power for the lasers
LD$1$ and LD$2$ was approximately $3$~mW and $4$~mW respectively, with
the laser beams focused inside the cell to diameters of about $200~\mu\mathrm{m}$.
\begin{figure} \centering
\includegraphics[width=1.0\columnwidth]{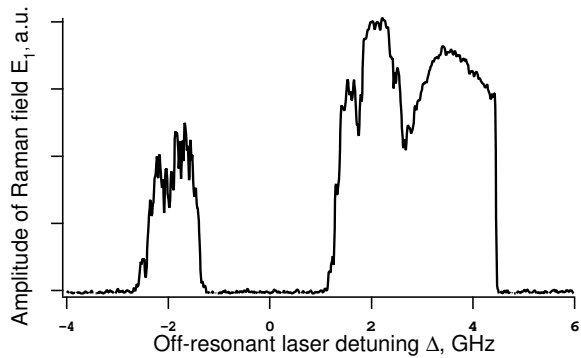}
\caption{Example spectrum of the generated Raman Stokes field $\mathrm{E}_1$ as
a function of the detuning $\Delta$ of the off-resonant pump field $\Omega_1$.
\label{raman_spec.fig}}
\end{figure}

We employed a cylindrical glass
cell filled with isotopically pure ${}^{87}$Rb
and $6$~Torr of Ne buffer gas. We placed the cell inside a three-layer
magnetic shield to reduce the influence of stray magnetic fields, and heated
the cell to $106^{o}$~C, which corresponds to a Rb vapor density of $7\cdot
10^{12} ~\mathrm{cm}^{-3}$.  Since counter-propagating laser beams can
produce mirrorless oscillations in the Raman field~\cite{zibrov'99prl} we
took care to avoid retro-reflection of either laser
beam back into the atomic cell.

For the chosen pump field polarizations and
resonant atomic levels, each of the generated Raman fields had circular
polarization, with the same chirality as the corresponding pump
field. Thus all fields had linear polarization after passing through the
second quarter-wave plate placed after the Rb cell, with the polarization of fields $\Omega_1$
and $\mathrm{E}_1$ being orthogonal to $\Omega_2$ and $\mathrm{E}_2$. All
these fields were then combined at the second polarizing
beamsplitter (PBS$2$) together with an additional beam from the first laser which
propagated outside of the atomic cell. The frequency of this ``bypass'' field
was shifted down by $80$~MHz with respect to the field $\Omega_1$. We
detected the beat-note signal between the bypass field and the Raman field
$\mathrm{E}_1$ using a fast photodetector (PD) and a spectrum
analyzer in zero-span mode with a registration bandwidth of $3$~MHz. Since
the amplitude of the bypass field was constant, only changes in the amplitude of
$\mathrm{E}_1$ were detected.
\begin{figure} \centering
\includegraphics[width=0.9\columnwidth]{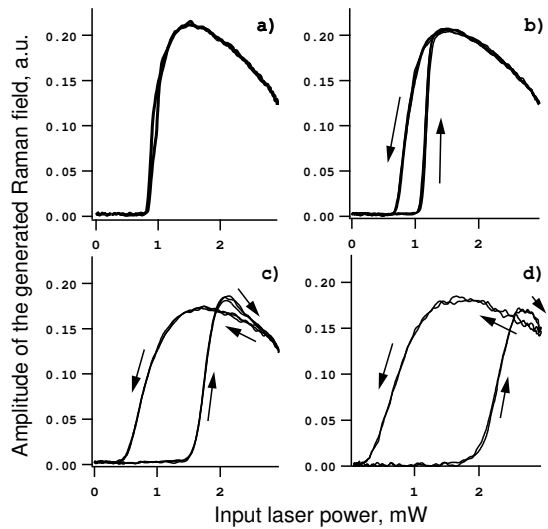}
\caption{ Differing degrees of observed hysteresis in the generated
Raman Stokes field $\mathrm{E}_1$ as the corresponding $\Omega_1$ pump field
intensity is modulated at a) $10$~Hz (no hysteresis); b) $150$~Hz; c)
$1.2$~kHz; d) $4$~kHz. Several ($2-3$) cycles are shown on each plot to
demonstrate the reproducibility of the data, with arrows indicating the
``direction'' of change of the pump field. The off-resonant laser detuning
$\Delta$ was approximately $4$~GHz.
\label{int_bist.fig}}
\end{figure}

Fig.~\ref{raman_spec.fig} shows a typical spectrum of the generated Stokes
field $\mathrm{E}_1$. We observed such Raman generation over a wide
range of the first laser's detuning $\Delta$. The absence of Raman
generation around $\Delta=0$ matches the resonant absorption
for the pump field $\Omega_1$. Raman generation also disappeared for
large detunings once the frequency of the generated field $\mathrm{E}_1$
approached the $F=1\rightarrow F'$ transitions. Other smaller variations of the
Stokes field amplitude were caused by effects such as switching between
different modes of Raman generation. The frequencies of these modes
differ by a few hundred kHz, and were found to depend on the intensity of the
pump fields as well as details of the laser beams' spatial overlap, which likely
affect the four-photon phase-matching
conditions~\cite{lukin'00,korsunsky'99pra}. However, these smaller Raman
variations did not change the qualitative behavior of the dynamic optical
bistability.

To study the dynamics of Raman generation,
we varied the intensity of the off-resonant laser field
$\Omega_1$ by modulating the voltage applied to the AOM at frequency
${f}_{mod}$, while the intensity and frequency of laser LD$2$ were kept
constant. Fig.~\ref{int_bist.fig}a shows the observed dependence of the
Stokes field amplitude $\mathrm{E}_1$ on the
corresponding pump field $\Omega_1$, for the case of very slow variations of
$\Omega_1$ (i.e., $f_{mod}T_2\ll1$). One can see that the amplitude of
the Stokes field exhibits threshold-like behavior, and then reaches a
maximum (determined by the phase matching conditions between
all four optical fields). If the intensity of the pump field increases
sufficiently, it drives the system out of the optimal conditions for
Raman generation, and the amplitude of the generated field decreases. It is
important to note that in the quasi-static case (${f}_{mod}<10$~Hz) the Raman
generation threshold is the same regardless of the ``direction'' of the pump
field's change. Thus there is no hysteresis in the regime of low modulation
frequency.
\begin{figure}
\centering
\includegraphics[width=1.0\columnwidth]{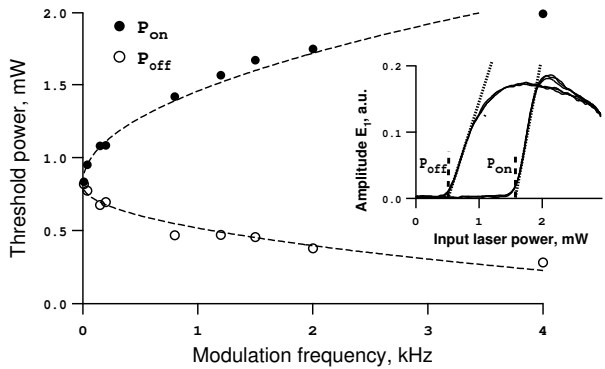}
\caption{ Measured dependence on modulation frequency of the
hysteresis threshold powers, $\mathrm{P_{on}}$ and $\mathrm{P_{off}}$, of the
off-resonant pump field. The dashed lines indicate independent fits of
$\mathrm{P_{on}}$ and $\mathrm{P_{off}}$ to $\sqrt{{f}_{mod}}$.
\textit{Inset}: Graphic definition of the manner in which $\mathrm{P_{on}}$ and
$\mathrm{P_{off}}$ were determined from the hysteresis data.
%are defined as zero-crossing points of the linear fits
%of the corresponding region of the hysteresis curve.
%
\label{threshold.fig}}
\end{figure}

As shown in Fig.~\ref{int_bist.fig}b-d, a hysteresis loop
appears for the Stokes field amplitude as $f_{mod}$ increases.
We define the ``threshold'' values $\mathrm{P_{on}}$ and $\mathrm{P_{off}}$
as the pump field powers at which Raman generation starts (for increasing
$\Omega_1$) and ceases (for decreasing
$\Omega_1$). Their dependence on $f_{mod}$ is
presented in Fig.~\ref{threshold.fig}.
%Note that these are for
%intensity modulation frequency, which is twice an AOM modulation
%frequency
One can see that the difference between $\mathrm{P_{on}}$ and $\mathrm{P_{off}}$
increases with the modulation frequency, and these threshold powers are reasonably fit
by $\mathrm{P_{on}}-\mathrm{P_{th}} \propto \sqrt{f_{mod}}$ and
$\mathrm{P_{off}}-\mathrm{P_{th}} \propto -\sqrt{f_{mod}}$, where
$\mathrm{P_{th}}$ is the threshold power for Raman generation in the
CW regime. We can reproduce this scaling law by taking into account the delayed response of the  generated Raman field to the pump field modulation, as characterized by the delay time $\tau_S$ (which depends on pump field power, see Eq.(\ref{tau_s})). This simple model yields:
\begin{equation} \label{threshold}
\mathrm{P_{on,off}} \simeq \mathrm{P_{th}}\left (1 \pm \frac12 \sqrt{f_{mod}\tau_S^{th}}\right)
\end{equation}
where $\tau_S^{th}$ is the four-photon dark-state equilibration time in the vicinity of the Raman generation threshold. This expression is in good qualitative agreement with the experimental results. However, a more complete description of the threshold behavior will require treatment of the diffusion of coherently prepared atoms in and out of the optical
interaction region~\cite{zibrov'01ol,zibrov'02pra}. 
\begin{figure}
\centering
\includegraphics[width=0.9\columnwidth]{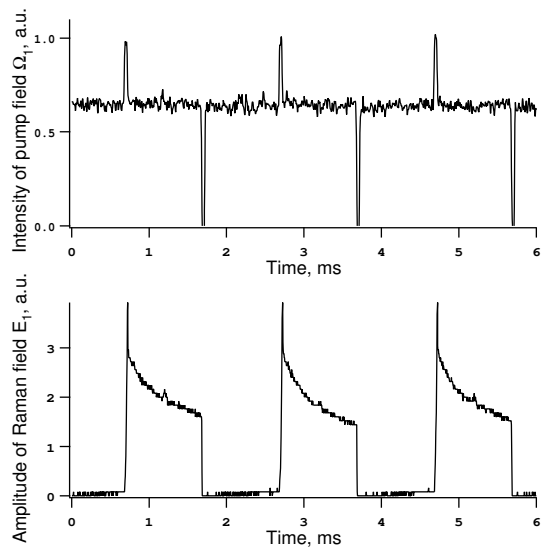}
\caption{Demonstration of an all-optical switch using dynamic optical
bistability. The Stokes field
$\mathrm{E}_1$ could be switched on and off (lower graph) by pulsing the
pump field $\Omega_1$ (top graph). The duration of the $\Omega_1$
pulses was $40~\mu\mathrm{s}$; the plotted intensity of these pulses is
normalized to their maximum value ($\approx 2.5$~mW) with a detuning $\Delta
\approx 4$~GHz.
\label{switch.fig}}
\end{figure}
%

%
%To explain this $\sqrt{{f}_{mod}}$ scaling, we recall that the atomic
%medium's response to a modulation P($t$) of the pump field power is delayed by
%$\tau_R$, given by Eq.(\ref{tau_ram}). Therefore, at low modulation frequencies,
%$f_{mod}\tau_R \ll 1$, turn-on or turn-off of Raman generation occurs at a
%shifted pump power $\tilde{\mathrm P}(t)=\mathrm P(t)-\tau_R\frac{d\mathrm P}{dt}$. Taking into
%account that $\tau_R$ is also modulated, $\tau_R \propto 1/|\Omega_1|
%\propto 1/\sqrt{\mathrm P(t)}$, and that near threshold $\mathrm P(t),\frac{d\mathrm P}{dt} \propto
%f_{mod}$, for low modulation frequencies, we find that the shifted threshold
%power is $\tau_R\frac{d\mathrm P}{dt} \propto \sqrt{{f}_{mod}}$.

At higher modulation frequencies (${f}_{mod}>1$~kHz) the measured peaks of the
Raman field $\mathrm{E}_1$ output differ for increasing and decreasing 
pump field $\Omega_1$, a result of changes in the four-photon phase-matching
conditions (which determine optimal Raman generation) that are fast relative
to $\tau_S$.
%
%We determined this coherence lifetime, in the presence of the optical fields, to
%be approximately $30~\mu\mathrm{s}$, as given by the half linewidth ($5$~kHz)
%of the beatnote signal between the pump field $\Omega_1$ and the Raman field
%$\mathrm{E}_1$. However, in a related experimental set-up
%we measured the ground-state coherence lifetime in the absence of optical
%fields to be much longer ($\sim 2$~ms). We speculate that the
%observed Raman hysteresis at modulation frequencies as low as $100$~Hz (see
%Fig.~\ref{int_bist.fig}b) arises from atoms that diffuse out of the optical
%interaction region and then return~\cite{zibrov'01ol,zibrov'02pra}. Further
%experimental study of this issue is needed.

Two additional notes about the observed dynamic optical bistability: (i) we
found Raman hysteresis similar to that shown on  Fig.~\ref{int_bist.fig} by
fixing the intensity of both pump lasers, tuning $\Delta$ to be near a cut-off
frequency for Raman generation (see Fig.~\ref{raman_spec.fig}), and modulating
the frequency of either of the pump fields; and (ii) we observed similar
bistable behavior for the anti-Stokes field $\mathrm{E}_2$, although this field
was generally much weaker than $\mathrm{E}_1$ because of large residual
absorption.
% It is known that interaction of the coherent electromagnetic fields with resonant
% media allows the manipulations of the one or several light fields by means of the
% other. In particular, there is an interest in realization of all-optical switches.
% Although large portion of the research is focused on the operation on a single-photon
% level~\cite{harris'98prl,harris'03}, there is still significant effort in studying the
% mechanisms which would allow to create a broadband all-optical switches and modulators
% for communication applications~\cite{hemmer'00prl,xiao'2003apl,hemmer'03prb}.

As an example application of dynamic optical bistability, we demonstrated
an all-optical switch in which the Raman field $\mathrm{E}_1$ was turned on
and off by briefly pulsing the intensity of the $\Omega_1$ pump field
above or below a median level (see Fig.~\ref{switch.fig}). During the turn-on
pulse, the strength of the pump field $\Omega_1$ was large enough to
establish significant atomic coherence, and thus to provide large Raman
generation once $\Omega_1$ returned to its median
value.
Similarly, the duration of the turn-off pulse was long enough for the resonant
$\Omega_2$ pump field to convert atomic coherence in the interaction region into
anti-Stokes $\mathrm{E}_2$ photons, in a process closely related to the
release of stored light~\cite{wal'03,kuzmich'03nature,hau'01nature,phillips'01prl,zibrov'02prl}. Thus
the switching for both turn-on and turn-off can likely be made much faster
with stronger pump fields.

In conclusion, we have studied the dynamics of resonantly-enhanced Raman
generation in a double-$\Lambda$ configuration in Rb vapor. We
observed a novel form of dynamic optical bistability based on long-lived atomic
coherence, which did not involve an optical cavity or an induced Bragg
grating in the medium. This bistability can be easily adjusted with changes to
the pump laser fields, which may assist practical applications. Realization of
dynamic optical bistability may also be possible in condensed matter systems~\cite{TurukhinSSMHH02,phillips'02,phillips'03}. As an example application we used dynamic
optical bistability to demonstrate a simple, all-optical switch.

We thank M.\ Lukin for the use of his laboratory, and M.\ D.\ Eisaman, L.\ Childress, V.\ A.\ Sautenkov, and P.\ Zoller for
useful and stimulating discussions. Financial support was provided by NSF grant
PHY-0113844, the Defense Advance Research Project Agency, the office of Naval
Research, NASA and the Smithsonian Institution.

\end{document}